\documentclass[sigconf]{acmart}

\AtBeginDocument{%
  \providecommand\BibTeX{{%
    \normalfont B\kern-0.5em{\scshape i\kern-0.25em b}\kern-0.8em\TeX}}}

\setcopyright{acmcopyright}
\copyrightyear{2023}
\acmYear{2023}
\setcopyright{acmlicensed}\acmConference[ASPDAC '23]{28th Asia and South
Pacific Design Automation Conference}{January 16--19, 2023}{Tokyo, Japan}
\acmBooktitle{28th Asia and South Pacific Design Automation Conference
(ASPDAC '23), January 16--19, 2023, Tokyo, Japan}
\acmPrice{15.00}
\acmDOI{10.1145/3566097.3567944}
\acmISBN{978-1-4503-9783-4/23/01}

\begin{document}

\title{A Multimode Hybrid Memristor-CMOS Prototyping Platform Supporting Digital and Analog Projects}

\author{K.-E. Harabi, C. Turck}
\author{M. Drouhin, A. Renaudineau}
\author{T. Bersani-{}-Veroni, D. Querlioz}
\email{damien.querlioz@c2n.upsaclay.fr}
\affiliation{%
  \institution{Univ. Paris-Saclay, CNRS}
  \city{Palaiseau}
  \country{France}
  \postcode{91120}
}

\author{T. Hirtzlin}
\author{E. Vianello}
\affiliation{%
  \institution{CEA-LETI, Univ. Grenoble-Alpes}
  \city{Grenoble}
  \country{France}
  \postcode{91120}
}

\author{M Bocquet}
\author{J.-M. Portal}
\affiliation{%
  \institution{Aix-Marseille Univ., CNRS}
  \city{Marseille}
  \country{France}
  \postcode{91120}
}

\renewcommand{\shortauthors}{Harabi et al.}

\begin{abstract}
We present an integrated circuit fabricated in a process co-integrating CMOS and hafnium-oxide memristor technology, which provides a prototyping platform for projects involving memristors. Our circuit includes the periphery circuitry for using memristors within digital circuits, as well as an analog mode with direct access to memristors. The platform allows optimizing the conditions for reading and writing memristors, as well as developing and testing innovative memristor-based neuromorphic concepts.
\vspace{-0.2cm}
\end{abstract}




\begin{teaserfigure}
\centering
  \includegraphics[width=1.0\textwidth]{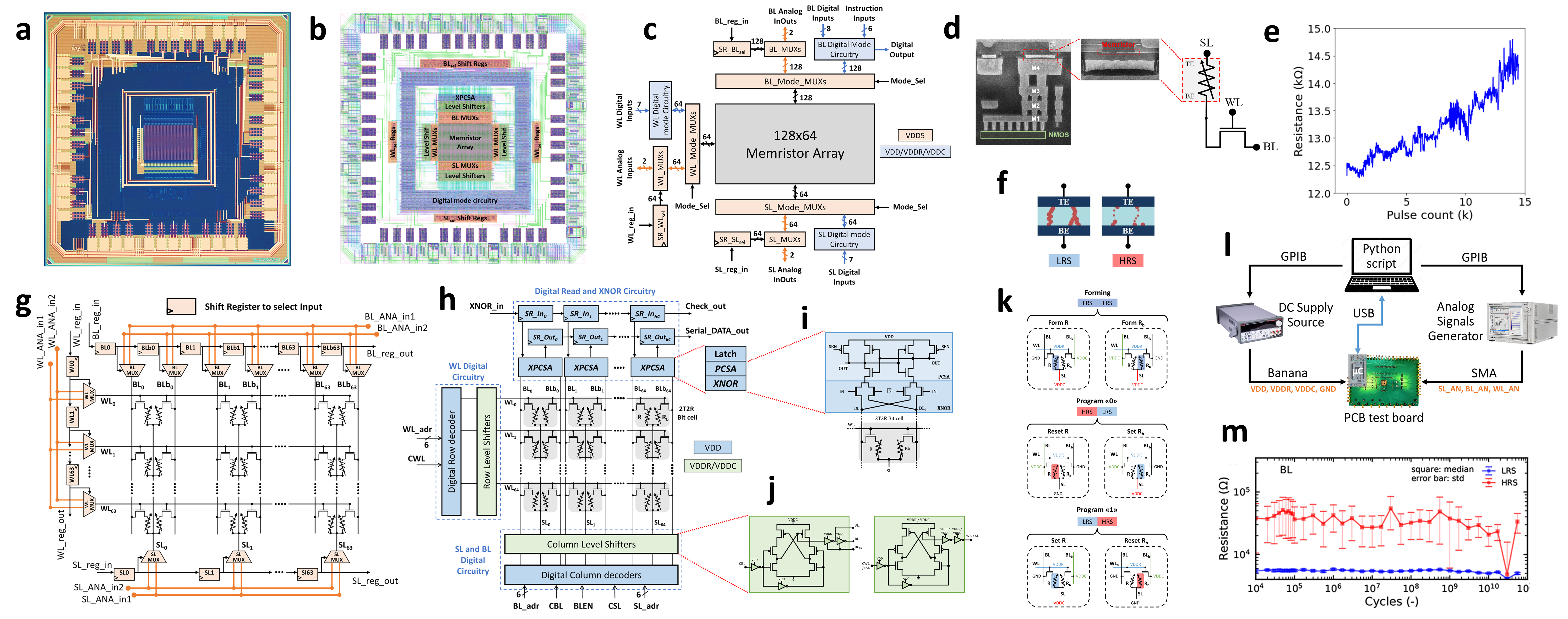}
  \caption{
\textbf{a}~Optical microscopy photograph, \textbf{b}~layout view, and \textbf{c}~ schematic  of the hybrid Memristor-CMOS die.
\textbf{d}~Electron microscopy image of a memristor in  our hybrid memristor/CMOS process.
\textbf{e}~Measurement of memristor resistance as a function of number of RESET programming pulses, for implementing a synaptic learning rule.
\textbf{f}~Illustration of memristor programming states. 
\textbf{g}~Schematic of the analog mode circuitry, with shift registers selecting inputs via Multiplexers .
\textbf{h}~Schematic of the digital mode circuitry, with a complementary 2T2R memristor basic cell.
\textbf{i} Schematics of the sensing circuitry with XNOR logic-in-memory feature. 
\textbf{j}~Schematic of the level shifters, used for shifting digital nominal voltage to forming and programming voltages of memristors.
\textbf{k}~Voltages applied for forming or programming a complementary cell in the digital mode.
\textbf{l}~Measurements setup of the prototyping platform.
\textbf{m}~Memristor endurance study, using the digital mode for programming and the analog mode for resistance measurements.}
  \Description{Hybrid Memristor-CMOS Prototyping platform.}
  \label{fig:teaser}
\end{teaserfigure}


\maketitle
\vspace{-0.5cm}
\section{Introduction}
Memristors, also known as resistive random access memories (RRAM) are a new type of memory technology fully embeddable in CMOS, providing a compact nonvolatile, and fast memory \cite{zidan2018future}.
These devices provide fantastic opportunities to integrate logic and memory tightly and allow low-power computing, in particular for Artificial Intelligence models and neuromorphic computing \cite{yu2018neuro}.
Unfortunately, the behavior of memristors is highly complex and partly stochastic \cite{majumdar2021model}: device models do not provide an accurate prediction of their behavior.
It is therefore essential to prototype computing concepts involving memristors experimentally.
However, appropriate platforms are extremely complex to fabricate due to the need to co-integrate commercial CMOS and  memristor devices on the same die. In this work, we designed, fabricated, and tested a prototyping platform, associating an array of 8,192 hafnium-oxide-based memristors and a collection of CMOS periphery circuits.
Our platform is multi-paradigm, which permits prototyping a wide range of both digital and analog projects.

\vspace{-0.25cm}
\section{Description of the Die}
A photograph and a layout view of our integrated circuit are presented in Figs.~1a-b. An electron microscopy image of a memristor in the backend of line of our hybrid memristor/CMOS process is shown in Fig.~1d. A commercial foundry fabricated the CMOS part (including the backend up to metal layer 4), using a 130-nanometer process. Afterward, we deposited the memristors on top of metal 4 using atomic layer deposition, and a fifth layer of metal. 

Our integrated circuit embeds periphery circuitry enabling the use of memristors within two modes. Fig.~1c shows a simplified schematic of the circuit. It uses consistent color codes: blue-colored blocks are digital-mode circuits, designed using thin oxide low-power transistors and supplied by digital nominal voltage (except for level shifters), and orange-colored blocks are analog-mode circuits, designed using thick oxide transistors to be compatible with high voltages.

The digital mode circuits (Figs.~1h-k) consist of: row and column decoders to select devices based on input addresses, level shifters on each row and column, which shift digital nominal voltage to higher voltages required to form and program memristors, and  precharge sense amplifiers, with a logic-in-memory feature, at each column  \cite{zhao2014synchronous}. These sense amplifiers allow reading the binary states of memory cells in a highly energy-efficient fashion while optionally performing logic operations at the same time. The complementary approach of \cite{bocquet2018} is used in our array for reducing the bit error rate.

When activating the analog mode (Figs.~1g), digital circuits are deactivated and the memristors array connections are switched to the analog circuitry. In this mode, shift registers configure input multiplexers permitting direct access to the analog state  of memristors, using low-resistance transmission gates. Each word line, bit line, and source line is then connected to the ground or to one of two analog InOut Pads, which can be connected to external equipment, e.g., Keysight B1530, a pulse source and measurement unit widely used to characterize memory devices.

The memristor array and all analog and mixed-signal circuits were designed in a full custom fashion, based on an extensive work on memristor characterization, modeling, and simulation. All digital circuits were placed and routed automatically using an HDL description and a Cadence Encounter flow provided by the foundry. Then, all circuits of the system were assembled manually and routed automatically using a Cadence Encounter flow developed in-house using a homemade abstract view of the memory array. 

\vspace{-0.25cm}
\section{Uses of the Platform}

To make the system re-configurable for different projects, we developed the  experimental setup of Fig.~1l: a PCB routes a microcontroller unit and  measurement equipment with our packaged die. Python scripts control the measurement. 

Optimizing read and programming strategies using the digital mode can allow the successful implementation of digital applications. Memristors feature a complex interplay between programming energy, reading speed, read disturb effects, and device endurance,  which our platform allows understanding. Fig.~1m shows an endurance study example. A memristor is programmed repeatedly using the digital mode circuits, and the memristor resistance is checked regularly using the analog mode and reported in Fig.~1m. It shows that the memristor resistance starts to degrade after $10^9$ cycles, concurrently with the emergence of bit errors seen by the reads after each programming using sense amplifiers (not shown in Figures). We observed that memristor endurance can vary between $10^3$ and $10^9$ cycles depending on programming conditions.

The analog mode of the platform can be used to prototype computing concepts where memristors are used in an analog fashion, e.g., as artificial synapses in machine learning or neuromorphic circuits \cite{yu2018neuro}. Fig.~1e shows measurements on a memristor in our platform when applying a succession of 15,000 1$V$ 1.5-$\mu$s programming pulses: the memristor resistance progressively increases, a feature that permits the memristor to implement a synaptic learning rule. This use is particularly appealing due to its compactness, but the imperfections of memristors (thermal and random telegraph noise,  cycle-to-cycle, and device-to-device variability) pose challenges that make it necessary to test ideas experimentally. Our platform supports prototyping various neuromorphic experiments, targeting inference, deterministic or probabilistic learning \cite{dalgaty2021situ}.

\vspace{-0.25cm}
\section{Conclusion}
We have designed, fabricated, and tested a flexible multi-paradigm platform to prototype and optimize digital and/or analog computing concepts, based on a hybrid CMOS/memristor integrated circuit.
We are currently using it to validate multiple digital logic-in-memory and analog neuromorphic concepts, and plan to make the platform available to other research groups.

\vspace{-0.25cm}

\begin{acks}
\vspace{-0.1cm}
This work was supported by  ERC Grant NANOINFER (715872) and ANR grant NEURONIC (ANR-18-CE24-0009).
\end{acks}
\vspace{-0.25cm}
\bibliographystyle{ACM-Reference-Format}
\bibliography{sample-base}


\begin{thebibliography}{6}


\ifx \showCODEN    \undefined \def \showCODEN     #1{\unskip}     \fi
\ifx \showDOI      \undefined \def \showDOI       #1{#1}\fi
\ifx \showISBNx    \undefined \def \showISBNx     #1{\unskip}     \fi
\ifx \showISBNxiii \undefined \def \showISBNxiii  #1{\unskip}     \fi
\ifx \showISSN     \undefined \def \showISSN      #1{\unskip}     \fi
\ifx \showLCCN     \undefined \def \showLCCN      #1{\unskip}     \fi
\ifx \shownote     \undefined \def \shownote      #1{#1}          \fi
\ifx \showarticletitle \undefined \def \showarticletitle #1{#1}   \fi
\ifx \showURL      \undefined \def \showURL       {\relax}        \fi
\providecommand\bibfield[2]{#2}
\providecommand\bibinfo[2]{#2}
\providecommand\natexlab[1]{#1}
\providecommand\showeprint[2][]{arXiv:#2}

\bibitem[Bocquet et~al\mbox{.}(2018)]%
        {bocquet2018}
\bibfield{author}{\bibinfo{person}{M. Bocquet} {et~al\mbox{.}}}
  \bibinfo{year}{2018}\natexlab{}.
\newblock \showarticletitle{In-Memory and Error-Immune Differential RRAM
  Implementation of Binarized Deep Neural Networks}. In
  \bibinfo{booktitle}{\emph{IEDM Tech. Dig.}} IEEE, \bibinfo{pages}{20.6.1}.
\newblock


\bibitem[Dalgaty et~al\mbox{.}(2021)]%
        {dalgaty2021situ}
\bibfield{author}{\bibinfo{person}{T. Dalgaty} {et~al\mbox{.}}}
  \bibinfo{year}{2021}\natexlab{}.
\newblock \showarticletitle{In situ learning using intrinsic memristor
  variability via Markov chain Monte Carlo sampling}.
\newblock \bibinfo{journal}{\emph{Nature Electronics}} \bibinfo{volume}{4},
  \bibinfo{number}{2} (\bibinfo{year}{2021}), \bibinfo{pages}{151--161}.
\newblock


\bibitem[Majumdar et~al\mbox{.}(2021)]%
        {majumdar2021model}
\bibfield{author}{\bibinfo{person}{A. Majumdar} {et~al\mbox{.}}}
  \bibinfo{year}{2021}\natexlab{}.
\newblock \showarticletitle{Model of the Weak Reset Process in HfO x Resistive
  Memory for Deep Learning Frameworks}.
\newblock \bibinfo{journal}{\emph{IEEE Trans. Elect. Dev.}}
  \bibinfo{volume}{68} (\bibinfo{year}{2021}), \bibinfo{pages}{4925}.
\newblock


\bibitem[Yu(2018)]%
        {yu2018neuro}
\bibfield{author}{\bibinfo{person}{S. Yu}.} \bibinfo{year}{2018}\natexlab{}.
\newblock \showarticletitle{Neuro-inspired computing with emerging nonvolatile
  memorys}.
\newblock \bibinfo{journal}{\emph{Proc. IEEE}} \bibinfo{volume}{106},
  \bibinfo{number}{2} (\bibinfo{year}{2018}), \bibinfo{pages}{260--285}.
\newblock


\bibitem[Zhao et~al\mbox{.}(2014)]%
        {zhao2014synchronous}
\bibfield{author}{\bibinfo{person}{W. Zhao} {et~al\mbox{.}}}
  \bibinfo{year}{2014}\natexlab{}.
\newblock \showarticletitle{Synchronous non-volatile logic gate design based on
  resistive switching memories}.
\newblock \bibinfo{journal}{\emph{IEEE TCAS I}} \bibinfo{volume}{61},
  \bibinfo{number}{2} (\bibinfo{year}{2014}), \bibinfo{pages}{443--454}.
\newblock


\bibitem[Zidan et~al\mbox{.}(2018)]%
        {zidan2018future}
\bibfield{author}{\bibinfo{person}{M~A Zidan}, \bibinfo{person}{J~P Strachan},
  {and} \bibinfo{person}{W~D Lu}.} \bibinfo{year}{2018}\natexlab{}.
\newblock \showarticletitle{The future of electronics based on memristive
  systems}.
\newblock \bibinfo{journal}{\emph{Nature electronics}} \bibinfo{volume}{1},
  \bibinfo{number}{1} (\bibinfo{year}{2018}), \bibinfo{pages}{22--29}.
\newblock


\end{thebibliography}

\end{document}